\documentclass[11pt]{article}
\pdfoutput=1
\usepackage{amsmath,amssymb}
\usepackage{graphicx}
\usepackage{xspace}
\usepackage{epstopdf}
\usepackage[left=1in, right=1in,bottom=1in]{geometry}
\usepackage{setspace}
\usepackage{footmisc}
\usepackage[font={footnotesize}]{caption}

\usepackage[round,numbers,sort&compress]{natbib}

\usepackage{mathptmx}

\date{}
\graphicspath{{./figs/}}

\newenvironment{lcase}
	{\left\lbrace \begin{aligned}}
	{\end{aligned}\right.}
\newcommand\DASH{Dam1\xspace} 
\newcommand\res{\ensuremath{\tau}} 
\newcommand\kb{k_{\text{break}}}
\newcommand\kz{k_{\text{unzip}}}
\newcommand\vps{v_\text{ps}}
\newcommand\vbb{v_\text{bb}}
\newcommand\nh{protofilament\xspace}
\newcommand\nsb{binding\xspace}
\newcommand\kt{\ensuremath{k_BT}\xspace}
\newcommand\pdev[2]{\ensuremath{\frac{\partial#1}{\partial#2}}}

\newcommand\dev[2]{\ensuremath{\frac{d#1}{d#2}}}
\newcommand\mtdam{\ensuremath{\Delta G_{\text{	\DASH}}}\xspace}

\newcommand\figref[1]{Fig.~\ref{#1}}
\newcommand\figrefl[2]{Fig.~\ref{#1}~{\it#2}}
\newcommand\eqnref[1]{Eq.~\ref{#1}}
\newcommand\mean[1]{{\langle{#1}\rangle}}

\newcommand\pdetach{\ensuremath{P_\text{detach}}}
\newcommand\tpf{t_\text{pf}}
\newcommand\tpfi{t_\text{pf,i}}
\newcommand\tzip{t_\text{unzip}}
\newcommand\Fpf{P_\text{pf}}
\newcommand\fpf{p_\text{pf}}
\newcommand\Fzip{P_\text{unzip}}
\newcommand\fzip{p_\text{unzip}}
\newcommand\Fnotzip{\bar{P}_\text{unzip}}
\newcommand\bb{burnt-bridge}
\newcommand\ps{powerstroke}
\newcommand\tp{\tilde t}
\newcommand\fe{f_\epsilon}

\title{Force transduction by the microtubule-bound \DASH ring}

\author{Jonathan~W~Armond \\
	Department of Physics and MOAC Doctoral Training Centre, \\
	University of Warwick, Coventry, CV4 7AL, UK
	\and Matthew~S~Turner\footnote{
     Corresponding author.  Address: 
     Department of Physics,
	   University of Warwick,
	   Gibbet Hill Road,
	   Coventry, CV4 7AL, UK,
	   m.s.turner@warwick.ac.uk} \\
	Department of Physics, \\
	University of Warwick, Coventry, CV4 7AL, UK}

\begin{document}
\maketitle

\begin{abstract}
  The coupling between the depolymerization of microtubules (MTs) and the motion
  of the \DASH ring complex is now thought to play an important role in the
  generation of forces during mitosis. Our current understanding of this motion
  is based on a number of detailed computational models. Although these models
  realize possible mechanisms for force transduction, they can be extended by
  variation of any of a large number of poorly measured parameters and there is
  no clear strategy for determining how they might be distinguished
  experimentally. Here we seek to identify and analyze two distinct mechanisms
  present in the computational models. In the first the splayed protofilaments
  at the end of the depolymerizing MT physically prevent the \DASH ring from
  falling off the end, in the other an attractive binding secures the ring to
  the microtubule. Based on this analysis, we discuss how to distinguish between
  competing models that seek to explain how the \DASH ring stays on the MT.  We
  propose novel experimental approaches that could resolve these models for the
  first time, either by changing the diffusion constant of the \DASH ring (e.g.,
  by tethering a long polymer to it) or by using a time varying load.

\end{abstract}
\singlespacing
\clearpage

\section*{Introduction}

Mitosis is the mechanism of cell division in eukaryotic cells. In mitosis,
chromosomes condense and are arranged at the center of the cell by the mitotic
spindle.  Microtubules (MTs) are protein fibers, composed of $n$ parallel
protofilaments (PFs, typically $n=13$) forming a hollow cylinder. Each PF is
built from stacked tubulin protein dimers. MTs emanate from centrosomes and
attach to chromosome-bound kinetochores. Centrosomes are positioned at both
poles of the cell forming a bipolar spindle. During anaphase chromosomes are
segregated and transported to the cell poles by the retraction of MTs, providing
both daughter cells with a single copy of the cell's chromosomes
\cite{Alberts2002}. To achieve segregation, depolymerizing kinetochore-attached
microtubules (KMTs) must generate forces, e.g., to overcome chromosomal drag in
the cytosol \cite{InouS1995}. There is evidence that mitotic MT force generation
occurs in the absence of MT minus-end directed motor proteins \cite{CoueLM1991}
and when minus-end depolymerization is inhibited \cite{GanemUC2005}. Previously,
a hypothetical sleeve had been proposed to couple MT depolymerization to
kinetochores \cite{KoshlMK1988,Hill1985}. A 10-protein complex, purified from
budding yeast \cite{CheesBWDAMYHDB2001}, called \DASH (or DASH) has been
observed to form rings around MTs \cite{MiranWSH2005,WesteAWNWDNB2005}. \DASH
rings have been observed tracking depolymerizing MT plus ends {\it in vitro}
\cite{WesteWADNB2006} and an optical trap has been used to measure
force-distance traces for \DASH-coated polystyrene beads attached to
depolymerizing MTs
\cite{AsburGPFD2006,FrancPGGDA2007,GrishEVSGWDBMA2008}. Intriguingly, \DASH has
been shown to be essential for chromosome segregation in budding yeast
\cite{JonesHGW2001,CheesEMDB2001} and important for avoiding mis-segregation
problems in fission yeast \cite{FrancMM2007}.

\subsection*{Mechanisms of force transduction}

Several models have been proposed to explain how the \DASH ring can couple the
kinetochore to a depolymerizing MT so as to produce a force. Over two decades
ago Hill proposed the first quantitative model describing how a depolymerizing
microtubule could be harnessed for the production of force \cite{Hill1985}. In
this model a hypothetical sleeve surrounds the MT and provides the attachment to
the kinetochore. An attraction between the sleeve and MT provides an energy
barrier preventing detachment, but this sleeve may still be able to slide along
the MT without paying the energy of detachment. More recent computational models
are detailed, mechanistic and micromechanical. One such model has taken into
account the energy predicted to be available due to the curling of PFs
\cite{MolodGEMA2005,MandeMM1991,Chr'FK1995} and, following the discovery of the
\DASH ring, was extended to reflect current structural knowledge and incorporate
the hypothesis that \DASH forms rigid transient links to the MT
\cite{EfremGMA2007}. Another independent model postulated an electrostatic
attraction maintaining the rings position at the tip of the MT \cite{LiuO2006},
combined with a powerstroke. All recent models include a combination of the
following features: 1) the intrinsic diffusion of the \DASH ring; 2) an
effective powerstroke due to curling PFs; 3) an attractive potential between
\DASH and the MT. However, whilst these models include many of the relevant
physical features of the system and produce satisfactory simulations of a
reliable force transduction system, the problem cannot be considered ``solved''
because many variants on these models are possible and they have not been
quantitatively compared to data. Furthermore, the lack of discriminatory
experimental data precludes validation. In light of this we feel that much can
be gained from rigorously analyzing the contribution of the various features in
order to determine their possible role.

In what follows we describe two distinct ``minimal'' models, both of which
describe a functional \DASH-mediated force transduction system. In the {\it \nh}
model the splaying PFs at the depolymerizing end physically prevent the ring
from sliding off. In {\it \nsb} models an attraction between the ring and MT
provides an energy barrier preventing detachment. The two models are not
mutually exclusive -- a hybrid model, incorporating both contributions, may also
apply although one of the constituent mechanisms will typically dominate. While
it is straightforward to modify our analysis to include such hybrid models we
neglect them here for clarity, since our purpose is to differentiate the
contributions. In common with previous models we also neglect other molecular
components, e.g., microtuble-associated proteins (MAPs) and kinases
\cite{CheesAJGKYCDB2002,ShimoGGFWEMRNYMBOD2006}, that certainly play important
additional roles {\it in vivo}.

Some previous studies have incorporated a powerstroke, arising from the motion
of PFs, in driving the motion of the \DASH ring
\cite{MolodGEMA2005,EfremGMA2007,LiuO2006}. Indeed, it has been demonstrated
that PFs can push a bead attached to the side of a MT \cite{GrishMAM2005}, with
a force of about 5~pN per 1--2 PFs. However, it is also known that models in the
``burnt bridges'' \cite{MorozPKA2007} class require no powerstroke {\it per se}
to generate motion \cite{ArtyoMK2008}. Rather, purely diffusive Brownian motion
can be rectified if ``bridges'' (here segments of MT) are lost (depolymerize)
after they have been crossed. No instantaneous {\it physical} force is required,
although the resultant rectified Brownian motion does give rise to a force in
the thermodynamic sense. Such models are, in turn, members of a larger class of
models known as ``Brownian ratchets'' \cite{PeskiOO1993}. These models exhibit
velocities that depend on applied force, and stall for sufficiently high forces,
as also seen in more complex models \cite{LiuO2006}. It is not clear {\it a
  priori} to what extent the powerstroke plays an important role. In the present
work we also seek to answer this question.


\subsection*{Generalized model}
We seek to analyze a general model that includes both a diffusive burnt bridge
mechanism and a powerstroke, in order to determine their relative
contribution. Here the powerstroke involves a depolymerization event which
``unzips'' PFs and moves the position of the last unbroken section of MT; a new
section takes on this identity when the previous one unzips (contains separated,
splayed protofilaments).  As a highly energetic powerstroke this is assumed to
occur even when the ring is very close to the MT end, with a rate that gives
rise to a depolymerization velocity $\vps$. The sequence of microscopic PF
unzippering events gives rise to a well defined velocity for the last fully
intact MT section, irrespective of the sequence in which the neighboring PFs
unzip, and the precise MT helicity. Critically, we also assume that polymerized
MT can be lost with a {\it second rate}, giving a depolymerization velocity
$\vbb$, whenever the ring has diffused a distance $\delta$ from the end. This
can be thought of as the depolymerization velocity of a bare MT because, in this
case, there is no \DASH ring anywhere on the MT. We make no prior assumptions as
to which contribution dominates, rather we determine this by fitting the
parameters $\vbb$, $\vps$ and $\delta$ to data for the variation of the \DASH
velocity with load \cite{FrancPGGDA2007}.

\begin{figure}
\begin{center}
\includegraphics[width=3.25in]{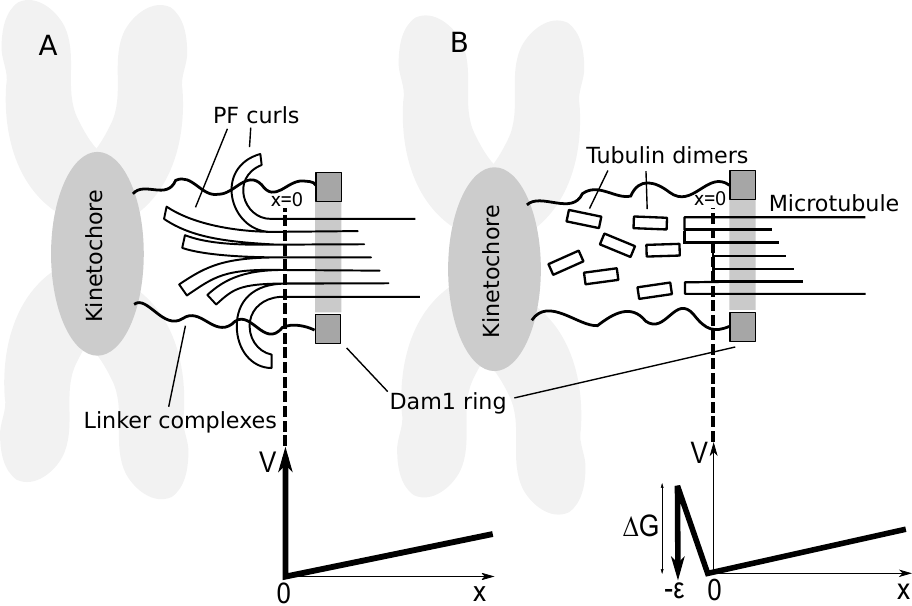}
\caption{Two general classes of models of \DASH ring-microtubule coupling. In
  both cases force is generated by rectifying Brownian motion, that is the ring
  diffuses to the right and the MT happens to unzip one segment or it is driven
  to the right by a powerstroke associated with unzipping. The dotted line
  denotes the point reached by MT unzipping ($x=0$). ({\it A}) The ring is
  sterically confined to the MT by PFs (\nh model) ({\it B}) The ring is
  attractively bound to the MT surface with a free energy of binding $\mtdam$
  (\nsb model). Below each model, the potential profile $V$ in which the ring
  diffuses is shown as a function of the distance $x$ of the ring from the MT
  end, and a dotted line indicates the connection between profile and model. The
  load force is the slope of $V(x)$ for $x>0$. In ({\it A}) there is a large
  (infinite) energy barrier preventing the \DASH ring moving to $x<0$ whenever
  curled PFs are present. If the PFs completely depolymerize, leaving a `blunt'
  end on the MT, this barrier disappears. In ({\it B}) the ring maintains only
  partial contact as it slides off the end of the MT ($-\epsilon<x<0$), which
  results in a rise in energy until it finally loses contact and is lost forever
  for $x<-\epsilon$. See text for details.}
\label{fig:models}
\end{center}
\end{figure}

Our model involves a clear distinction between two mechanisms (only) and
represents the simplest possible model capable of explaining this data. It can
be biophysically motivated on the grounds that the \DASH ring interacts with
neighboring tubulin and so the rate of PF unzippering at the MT end should
depend on how close the \DASH ring is to the end (a concept already introduced
in \cite{LiuO2006}). In the section below we discuss this mechanism in terms of
a putative energy landscape for the depolymerization (unzipping) reaction. We
believe that it would be unjustified to postulate the existence of any features
on this energy landscape beyond the minimum required to explain the data. This
amounts to a model involving two (distinct) depolymerization mechanisms. Our
results suggest that both mechanisms play important roles at moderate loads. In
particular \ps-only models are clearly inconsistent with the data showing a
velocity that is dependent on force \cite{AsburGPFD2006,FrancPGGDA2007},
assuming a strong power-stroke as previously measured on the order of $30-65$~pN
\cite{GrishMAM2005}. This is because, with a \ps-only model, we would not expect
the velocity of \DASH ring to be significantly slowed under a force as low as
$2$~pN; the data shows a significant slowing.  We find that the length scale
$\delta$ controlling burnt-bridge reactions, a free fit parameter, is close to
the axial length of a tubulin dimer. We speculate that this may provide
indication of ``crack"-like splitting of the MT, as discussed below.

\section*{Model}
The \DASH ring complex is reported to be capable of axial movement with respect
to the MT \cite{WesteWADNB2006}. Therefore, we treat the \DASH ring as a
particle undergoing one-dimensional Brownian motion in a potential $V(x)$ (shown
for two different models in \figref{fig:models}). The fully intact MT extends
away from the depolymerizing end for $x>0$ and the point at which the MT lattice
unravels is $x=0$ (see \figrefl{fig:models}{A}). The following Fokker-Plank
equation \cite{Risken1984} determines the probability density $\phi(x,t)$ for
the ring's position relative to the (moving) end,
\begin{equation}\label{eq:fpe}
\pdev{\phi}{t} = D\pdev{}{x}\left( \pdev{\phi}{x} + \frac{1}{\kt}\pdev{V}{x} \phi\right),
\end{equation}
where $D$ is the diffusion constant of the ring. This approach is appropriate
providing the depolymerization velocity $v$ of the MT is not too fast
(bounds given later in this section), otherwise we must instead
treat this as a full moving boundary problem. Since the microtubule
depolymerization is here quasistatically slow with respect to the diffusive
relaxation of the ring, we can neglect the drag force on the ring, except as
discussed in Supporting Material.

In the following we assume the \DASH ring is sufficiently stable that it can
only dissociate by slipping off the tip. For simplicity we restrict our analysis
to continuous depolymerization processes only and discount the possibility of
rescue and polymerization. Although it would be straightforward to include such
processes we believe that they would distract from the central results of this
paper.

A force $-\partial V/\partial x$ appears in \eqnref{eq:fpe}. This is the
magnitude of the applied force $f$ on the \DASH ring while on the MT ($x>0$) since the
ring must do work to move against this force. Hence, from \eqnref{eq:fpe} it can
be shown that, for constant (or slowly varying) $f$ the probability distribution
$\phi(x)$ is of Boltzmann form
\begin{equation}\label{eq:boltzmann}
\phi(x) = \frac{f}{\kt} \exp\left( -\frac{fx}{\kt} \right),
\end{equation}
where the ring typically explores a characteristic diffusion length
$\lambda=\kt/f$ from the MT end and positive values of $f$ here indicate loads
pulling in the negative $x$ direction (towards the MT end). We assume that the
depolymerization is {\it quasistatically slow}. This is appropriate provided the
time for the MT to depolymerize the distance $\lambda$ is much larger than the
relaxation time for a ring to diffuse this distance. This in turn requires
$\lambda/v(f)\gg\lambda^2/D$. In this case the distribution of the ring position
is always close to the equilibrium probability distribution that it would have
on an MT that was not depolymerizing. This sets an upper bound on the
depolymerization velocity, or equivalently a lower bound on the load force,
beyond which our theory is at best semiquantitative; solving
  $D/v(f)=\kt/f$ with \eqnref{eq:velocity}, we estimate these values to
be 500~nm/s and 0.04~pN respectively. Under these conditions the average ring
velocity is equivalent to the depolymerization velocity of the MT.

\subsection*{Force dependent depolymerization velocity}

\begin{figure}
\begin{center}
\includegraphics[width=3.25in]{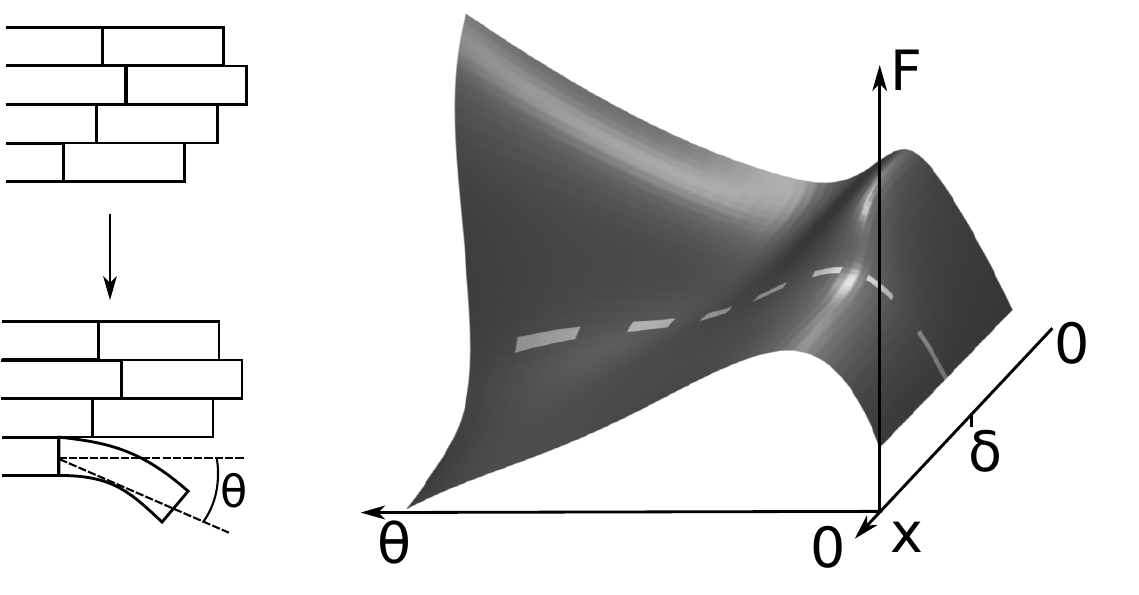}
\caption{Schematic energy landscape underlying PF unzipping. The proposed free
  energy $F$ landscape of a tubulin dimer at the end of the MT is shown (right)
  as a function of the distance of the \DASH ring from the MT end, $x$, and a
  reaction coordinate for the unzippering, the angle $\theta$ moved by the
  tubulin dimer (see diagram at left). The diagram is shown for illustrative
  purposes only and is not quantified in this work. Here $\theta=0$ represents a
  dimer in a linear PF incorporated into a stable MT. During unzippering
  $\theta$ increases and the dimer moves out, ultimately forming the base of a
  splayed PF. The unzippering is an activated process with an energy barrier
  (the height of the ridge on the right) that is {\it different} for a
  powerstroke ($x<\delta$) and a burnt bridges reaction ($x>\delta$), leading to
  velocities $\vps$ and $\vbb$ respectively. The energy landscape must have at
  least these basic features in order to give rise to the two depolymerization
  rates consistent with the data.}
\label{fig:energy}
\end{center}
\end{figure}

The \ps\ and \bb\ reactions can be thought of as arising from transitions over
an energy barrier of the form shown in \figref{fig:energy}, where the free
energy $F$ of PF curling is shown as varying with protofilament angle $\theta$
and the distance of the \DASH ring from the MT tip $x$. The figure shows only a
putative schematic of the free energy of PF curling reaction, and should not be
confused with the potential $V(x)$ in which the \DASH ring diffuses.

PFs may produce a power stroke that pushes the ring with force $f_\text{pf}$,
estimated from experimental evidence to be $30-65$~pN \cite{GrishMAM2005}. This
is the slope down the descending valley, diagonally right to left, in
\figref{fig:energy}. Provided that the load force $f\ll f_\text{pf}$ the \ps\
will give rise to a depolymerization velocity $\vps$ that is the rate at which
the last intact dimer on the MT crosses the highest part of the ridge-like
energy barrier in \figref{fig:energy} ($x<\delta$). Since the estimate for
$f_\text{pf}$ is so much larger than any force considered here, it is reasonable
to make the limited assumption that $\vps$ is constant for all experimentally
measurable load forces of a few pN or less.

In addition the MT can also depolymerize when the \DASH ring is further than a
critical distance $\delta$ from the end of the MT. In this case the \bb\
reaction gives rise to a depolymerization velocity $\vbb$ that is the rate at
which the last intact dimer on the MT crosses the {\it lower} part of the
ridge-like energy barrier in \figref{fig:energy} ($x>\delta$).  That the rate of
MT unzippering is {\it retarded} when the \DASH ring is near the MT end is a
result of the fact that the velocity decreases as the load force is increased
and the ring is more often closer to the MT end. Although it is not {\it
  necessary} to interpret our model in terms of the \DASH ring physically
occluding the unzippering of the tubulin dimers, this interpretation may not be
unreasonable, particularly in view of the fact that we find $\delta$ to be
comparable with the axial length of the last intact ring of tubulin dimers.

The resultant velocity due to both mechanisms is the sum of the probability that the ring is close to the MT end $x<\delta$, multiplied by the \ps\ velocity, and the probability that it is far $x>\delta$, multiplied by the \bb\ velocity,
\begin{align}
v &= \vps\left(1-\int_\delta^\infty \phi(x)\,dx \right) + \vbb \int_\delta^\infty \phi(x)\,dx \nonumber \\
&= \left( \vbb-\vps \right) \int_\delta^\infty \phi(x)\,dx + \vps
\label{eq:velone}
\end{align}

The velocity follows from Eqs.~\ref{eq:velone} and \ref{eq:boltzmann}
\begin{equation}\label{eq:velocity}
v = (\vbb-\vps) \exp \left( \frac{-f\delta}{\kt} \right) + \vps,
\end{equation}

\begin{figure}
\begin{center}
\includegraphics[width=3.25in]{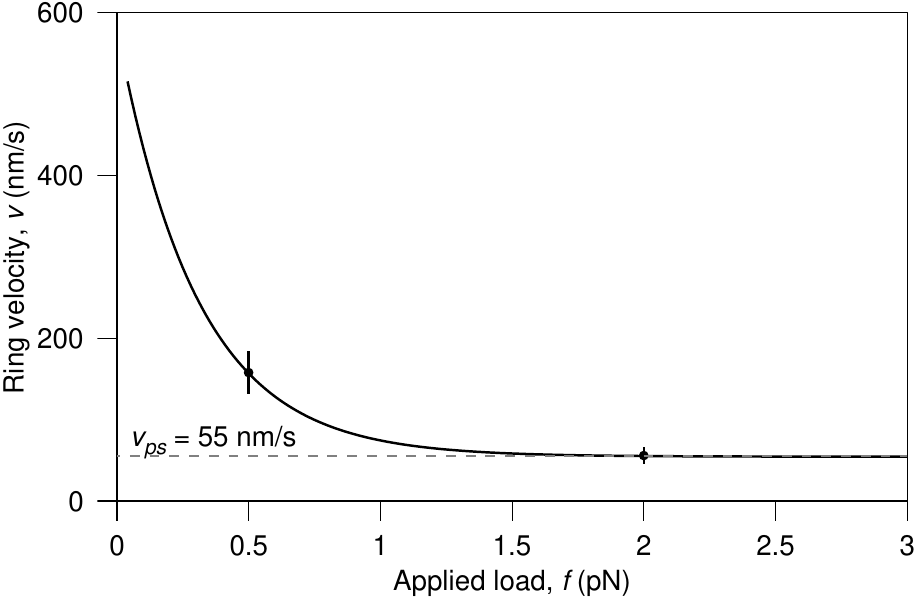}
\caption{The variation of velocity of the \DASH ring with applied load. The
  velocity falls as the force increases because the motion must increasingly
  rely on the energetic powerstroke. Note that, although the graph appears to
  suggest an absence of a stalling force, a significantly higher forces the
  assumption of constant $\vps$ would fail and the ring would stall. The curve
  is produced from the best fit of $\delta$ and $\vps$ in \eqnref{eq:velocity}
  and data from \cite{FrancPGGDA2007}.}
\label{fig:load}
\end{center}
\end{figure}

The variation of this velocity with load is shown in \figref{fig:load} for
$\vbb=580$~nm/s \cite{HuntM1998}, and the values $\vps=55$~nm/s and
$\delta=14$~nm that correspond to the best fit to data
\cite{FrancPGGDA2007}. Since a ``burnt-bridges''-only model fails to fit the
data sufficiently (i.e. $\vps>0$) it suggests that a powerstroke plays a role in
forced Dam1 motion. It should be noted that, although in this model
$v\rightarrow\vps$ as $f\rightarrow\infty$, we do not suggest this is a physical
feature of the system. Rather it is the consequence of the assumption that
protofilaments are perfectly rigid and the \ps\ reaction is asymptotically
strong. Our model would need modification for forces approaching
$f_\text{pf}$. As discussed later, PFs are estimated to require tens of pN to
bend.

\subsection*{Two models for \DASH ring retention}

\begin{figure}
\begin{center}
\includegraphics[width=5in]{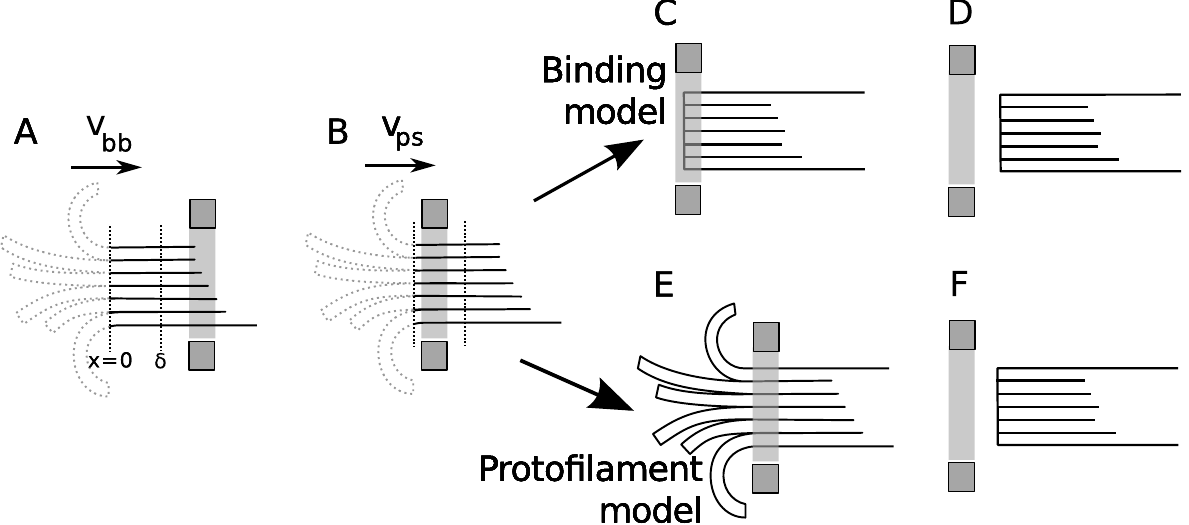}
\caption{Various sketches of a ring on a microtubule. ({\it A}) In this
  configuration the ring is further than $\delta$ from the tip of the MT, so the
  MT depolymerizes with velocity $\vbb$. Unzipped protofilaments are shown
  dotted as they do not affect depolymerization. ({\it B}) In some other
  configuration, the ring is closer to the tip than $\delta$, so the MT
  depolymerizes with velocity $\vps$. In ({\it C-F}) the detachment mechanisms
  are shown. This is either insensitive to PFs ({\it C-D}; binding model) or
  sensitive to PFs ({\it E-F}; protofilament model). In ({\it C}) and ({\it E})
  the ring has not yet escaped. In ({\it D}) and ({\it F}) the ring has escaped
  from the MT.}
\label{fig:timeline}
\end{center}
\end{figure}

We now proceed to calculate the mean time the \DASH ring will remain on a MT and
transduce force\footnote[1]{Recently it has been discovered that the \DASH
  oligomers track the tip of depolymerizing MTs without forming a ring
  \cite{GestaGCWZWAD2008,GrishSVEWDBAM2008}. It seems unlikely that a \nh model
  could operate without a full ring, however, it is not known to what extent, if
  at all, small oligomers contribute to force production. Furthermore, it has
  been shown that 16-20 \DASH complexes are present at the kinetochore during
  metaphase \cite{JogleBFLWBHSB2008}, enough to form the ring. We await the
  result of experiments where tension is applied to putative \DASH oligomers.
}. This time is controlled by different physics in the \nh\ and the \nsb\
models, see Figs. \ref{fig:models} and \ref{fig:timeline}. However, in both
cases, the {\it velocity} of the ring is governed by the model described above,
see \eqnref{eq:velocity}.

\subsubsection*{Runtime: Binding model}

The binding model involves a ring diffusing on a MT according to
\eqnref{eq:fpe}, leading to a depolymerization velocity as given in
\eqnref{eq:velocity}. However, in order to detach from the MT end the ring must
overcome a linear potential imposed by the \DASH-MT binding energy \mtdam as it
slides off the end of the ring. In this respect it is similar to Hill's model
\cite{Hill1985}. Previous models invoked a \mtdam that also determined the
roughness of the energy landscape through ``linkers'' \cite{EfremGMA2007} whose
existence is supported by binding studies \cite{MiranKH2007}. Here we don't make
this assumption, rather \mtdam could be due to less specific interactions
without significant energy barriers between neighboring sites
\cite{WangRWLWNDBN2007} but, importantly, can vary independently of the
diffusion constant $D$. This, in turn, is fixed by the smoothness of the
underlying energy landscape experienced by the ring as it diffuses along the MT
(distinct from the energy landscape experienced by an unzippering PF shown in
\figref{fig:energy}). This model assumes that the splayed PFs play no role,
either because they are transient (rapidly breaking) or otherwise interact
negligibly with the ring as it slides off the end of the MT.  Although clearly
an extreme approximation it forms the natural opposite limit to the \nh model
discussed in the next section. Under a load force the ring is in the well of a
``tick''-shaped potential with two linear domains (inset
\figrefl{fig:models}{B}). To move to the left (towards negative $x$) it must
partially unbind from the MT, to move to the right (positive $x$) it must do
work against the applied force. The potential gradients experienced by the \DASH
ring determine the load force $f$ (while on the MT, $x>0$) and the resultant
force $\fe$ (while detaching from the MT over the small distance
$-\epsilon<x<0$). The force on the ring, adopting a sign convention where a
positive force acts in the direction of positive $x$, is therefore given by
\begin{align} \label{eq:potential}
-\pdev{V}{x} =
 \begin{lcase}
   &-f &\quad x \geq 0 \\
   &\fe = \frac{\mtdam}{\epsilon} - f &\quad -\epsilon\leq x\leq 0 
 \end{lcase}
\end{align}
where $\epsilon$ is the unbinding region. If the ring is in the region
$x<-\epsilon$ then it is lost, and if lost we assume it never returns, hence we
have $V\to-\infty$ for $x<-\epsilon$.

Symmetry from electron microscopy \cite{WesteWADNB2006} and copy number \cite{JogleBFLWBHSB2008} experiments suggest 16 complexes are required to form the \DASH ring, however, the total bond energy may not be additive and this should therefore be regarded as an extreme upper bound on the total binding energy.

The detachment of the ring can be cast as a classical Kramers escape problem
\cite{Krame1940}. To solve \eqnref{eq:fpe} with \eqnref{eq:potential} we
followed the method in \cite{AgudoM1993,MalakP2002}. In this way we
obtain the lifetime of the metastable state directly from the Laplace
transformed version of \eqnref{eq:fpe}, with initial condition
$\phi(x,0)=\delta(x)$, where $\delta(x)$ is the Dirac delta function, although
the precise form of this initial condition is unimportant. The mean time the
ring remains on the MT is the runtime \res
\begin{equation}\label{eq:nsb_res}
 \res = \frac{(\kt)^2}{D\fe} \left(\frac{e^{\frac{\fe\epsilon}\kt}-1}{f} -
   \frac{e^{\frac{\fe\epsilon}\kt}+\frac{\fe\epsilon}{\kt}-1}{\fe} \right),
\end{equation}
where $\epsilon$ is a small distance.

\subsubsection*{Runtime: Protofilament model}

The \nh\ model involves a ring diffusing on a MT according to \eqnref{eq:fpe},
leading to a depolymerization velocity \eqnref{eq:velocity}, as before. However,
in order to detach from the MT end the ring has to wait until all protofilaments
have broken (depolymerized), leaving a sufficiently ``blunt'' end to the MT for
the ring to simply slide off, see \figrefl{fig:models}{A}. We no longer require
the \DASH ring to overcome a \DASH-MT binding energy. Electron microscopy
reveals that short, separated PFs splay outwards at the depolymerizing MT end
\cite{MandeMM1991,Chr'FK1995} and it is quite plausible that these block the
escape of the ring; the elastic energy required to straighten a curled PF
\cite{Landau1986} follows from measurements of their rigidity
\cite{KikumKTT2006,GitteMNH1993} and is of the order of tens of $\kt$ per
subunit, i.e. very large.

The ``frayed'' PFs near the end of the MT are curved and laterally separate. The
unzipping (depolymerization) of the MT lattice (see \figref{fig:models}A) is
most accurately described as a process which transfers length from the
polymerized MT into separated PFs.  The unzipping is thought to be driven by the
stored elastic energy in the $\alpha\beta$-tubulin units in the lattice
\cite{RiceMA2008}. When not constrained by lateral bonds, PFs relax into a
curved state. We model unzipping as a Poisson process with rate $\kz$. Each
unzipping event extends every PF curl by some microscopic, or subunit, length
$b$, leading to a depolymerization velocity $v=b\kz$. This microscopic length
might be the tubulin dimer repeat distance $b$, if the MT splits between a
particular pair of PFs, or otherwise smaller than this. When the ring is within
a small length $\delta$ unzipping is inhibited. To more carefully analyze this
process note that the time between unzipping events $\tzip$ is an exponential
random variable, with probability density function $\fzip(t)=\kz \exp(-\kz t)$,
and mean
\begin{equation}\label{eq:t_zip}
\mean{\tzip} = \frac{1}{\kz} = \frac{b}{v}.
\end{equation}

The distribution of the ring position in this model follows
\eqnref{eq:boltzmann}. Detachment occurs when all PF curl lengths reach
zero\footnote[2]{Extensions of our model to the case of loosely-fitting rings is
  straightforward, involving attachment whenever the PF curls exceed some finite
  length $L$. Our results are qualitatively insensitive to this modification,
  provided the ring rarely detaches at low force. Furthermore, for such rings
  the molecular length $b$ becomes irrelevant as the characteristic timescale is
  $L/v$.}. Since $v$ is a function of the applied force $f$, according to
\eqnref{eq:velocity}, $\mean{\tzip}$ increases under load. From
\eqnref{eq:boltzmann}, we have that the characteristic distance of the ring from
the tip is $\lambda=\kt/f$. The characteristic time for the ring to diffuse this
length and escape is $\lambda^2/D$ which is much less than $\mean{\tzip}$ for
typical parameters whenever $f>0.15$~pN (see {\it Supporting Material}). Thus it
is not unreasonable to assume that the ring might disengage from the MT
extremely rapidly as soon as all curled PFs reach zero length.

We assume that tubulin subunits on the frayed PFs break independently according
to a Poisson process with rate $\kb$. The depolymerization of PFs then follows
from the loss of all PF material beyond the break, as in previous computational
models \cite{VanBuOC2002}. A PF curl reaches zero length if the axial bond
nearest to the unzipping point breaks, see \figrefl{fig:models}{A}. Since this
occurs with a rate $\kb$ the waiting time $\tpfi$ for PF curl $i$ to break off
completely is an exponential random variable. The wait time for all $n$ PF curls
breaking is the order statistic $\tpf=\max_i \tpfi$. The distribution function
for this time is $\Fpf(t) = (1-\exp(-\kb t))^n$ and the mean wait time (see
section 4.6 from \cite{Arnold1992} or \cite{R'en1953}) is

\begin{equation}\label{eq:t_break}
\mean{\tpf} = \sum^n_{i=1} \frac{\mean{\tpfi}}{n-i+1} = \frac{H_n}{\kb},
\end{equation}
where $H_n=\sum^n_{i=1} i^{-1}$ is the harmonic number, roughly $\log n$ for
$n\gg 1$ as can be seen by converting the sum to an integral, $\mean{\cdot}$
denotes the ensemble average. The \DASH ring will therefore no longer be secured
to the MT end and will detach after a time $\tpf$ provided that no unzipping
events having taken place during the time $\tpf$. If the MT has unzipped then
the PFs extend (from their base), effectively ``restarting'' the waiting
process.

Fundamentally we are interested in the mean runtime $\res$, this being the time
taken for the curled PFs to all depolymerize completely even while the MT is
simultaneously undergoing stochastic unzipping events. $\res$ can be found by
counting the number of unzipping events $N$ that occur before the PFs all
successfully break and the \DASH ring can disengage. $N$ is geometrically
distributed with mean $\mean{N}=1/\pdetach$ with $\pdetach$ the probability that
the curled PFs depolymerize completely before the next unzippering event. Thus
\begin{equation}\label{eq:nh_def}
 \res  = \frac{\mean{\tzip}}{\pdetach}.
\end{equation}

The ring detaches if the PFs break before an unzipping occurs, i.e. with probability that $\tpf < \tzip$,
\begin{equation}
 \pdetach = \int_0^\infty dt \, \fzip(t) \int_0^t \fpf(t^\prime)  \,dt^\prime,
\end{equation}
where $\fpf = d\Fpf/dt$ is the probability density function for $\tpf$. Evaluating the integral with respect to $t^\prime$
\begin{align}
 \pdetach &= \int^\infty_0 \Fpf(t) \fzip(t) \,dt \nonumber \\
          &= \int^\infty_0 \left( 1-e^{-\kb t} \right)^n \kz e^{-\kz t}\,dt,
\end{align}

Binomially expanding the integrand, integrating term-by-term and substituting
back into \eqnref{eq:nh_def} we obtain
\begin{equation}\label{eq:nh_sol}
\res = \frac{1}{\kz} \left( \sum^n_{j=0} \binom{n}{j}  \frac{\kz(-1)^j}{j \kb + \kz}\right)^{-1}.
\end{equation}
where $\binom{n}{j} = n!/j!(n-j)!$.

\subsection*{Time varying applied forces}
We now consider an oscillating applied force of the form
\begin{equation}\label{eq:time_varying_force}
f(t)=f_0\sin \omega t + f_1
\end{equation}
Provided the period is sufficiently long $\omega^{-1}\gg\lambda^2/D$ our
quasi-static approximation for the ring position should give an accurate
estimate for its probability density $\phi(x,t)$.

The depolymerization velocity will be retarded according to
\eqnref{eq:velocity}, relating $v$ to $f(t)$.

\subsubsection*{Protofilament model under oscillating force}
The probability that the MT does {\it not} unzip in a time $t$ after the time at which the last unzipping occurred $\tp$ is
\begin{equation}\label{eq:FZphi}
\Fnotzip(t;\tp) = 1-\Fzip(t;\tp) = \exp\left( -\int_{\tp}^{\tp+t} \frac{v(t^\prime)}{b}\,dt^\prime \right),
\end{equation}

Since \eqnref{eq:FZphi} depends explicitly on $\tp$, we perform an average over
$\tp$, appropriately weighted, to give the complementary distribution of times
between unzipping events
\begin{align}
\Fnotzip(t) &= \int_0^{2\pi/\omega} \Fnotzip(t;\tp) \frac{v(\tp)}{\mathcal{N}b} \,d\tp \nonumber \\
&= \int_0^{2\pi/\omega} \exp\left( -\int_{\tp}^{\tp+t} \frac{v(t^\prime)}{b}\,dt^\prime \right) \frac{v(\tp)}{\mathcal{N}b} \,d\tp
\end{align}
involving a normalization constant $\mathcal{N}=\int_0^{2\pi/\omega} \frac{v(\tp)}{b} \,d\tp$.

To calculate the runtime as in \eqnref{eq:nh_def}, we first determine the
probability the unzip time exceeds the curled PF breaking time
\begin{align}
\pdetach &= \int_0^\infty dt \, \fzip(t) \int_0^t \fpf(t^\prime) \,dt^\prime \nonumber\\
&= \int_0^\infty \,dt^\prime \fpf(t^\prime) \int_{t^\prime}^\infty \fzip(t) \,dt \nonumber\\
&= \int_0^\infty \Fnotzip(t) \fpf(t)  \,dt.
\end{align}

The probability density of $\tpf$ is
\begin{align}
\fpf(t) &= \dev{}{t}\Fpf(t) = \dev{}{t}\left( 1-e^{-\kb t}\right)^n \nonumber\\
&= n\kb e^{-\kb t}\left(1-e^{-\kb t}\right)^{n-1}.
\end{align}

Finally the runtime is
\begin{align}\label{eq:nh_res_osc}
\res &= \left[ \pdetach \mean{\kz(t)} \right]^{-1} \nonumber \\
&= \frac{1}{\pdetach} \frac{\omega}{2\pi} \int_0^{2\pi/\omega} \frac{b}{v(t)} \,dt,
\end{align}
where $1/\pdetach$ is the mean number of steps before detachment.

\subsubsection*{Binding model under oscillating force}
The generalization of \eqnref{eq:nsb_res} to the  case of time-varying force (\eqnref{eq:time_varying_force}) is straightforward,
 
\begin{equation}\label{eq:nsb_res_osc}
 \res = \frac{\omega}{2\pi} \int_0^{2\pi/\omega} \frac{(\kt)^2}{D\fe} \left(
   \frac{e^{\fe\epsilon/\kt}-1}{f} - \frac{e^{\fe\epsilon/\kt} +
     \fe\epsilon/\kt -1}{\fe} \right)  \,dt,
\end{equation}
where $f$ and $\fe$ are now {\it time-dependent} potential gradients, according
to \eqnref{eq:time_varying_force} with \eqnref{eq:potential}.

\section*{Results and Discussion}

We identified the following parameters using data reported in the experimental
literature; $\vbb=580$~nm/s \cite{HuntM1998},
$D=0.083\pm0.001$~$\mu\text{m}^2$~$\text{s}^{-1}$ \cite{WesteWADNB2006}; for the
\nh model we assume $b=8$~nm and $n=13$ to be typical.

Table 1 in \citet{FrancPGGDA2007} (see Table S1 in the Supporting Material)
lists velocities at $f=0.5$~pN and $2.0$~pN. Using the velocity data we fit to
obtain $\delta=14\pm1.4$~nm and $\vps=55\pm9.3$~nm/s. As has already been
mentioned the range $\delta$ is intriguingly close to a tubulin axial repeat
length ($8$~nm or 1.5 times this, due to helicity). A simple picture might be of
a sleeve that suppresses depolymerization while it sits over the next intact
tubulin dimers in the PFs that are about to split. This supports the idea that
the MT depolymerizes by first splitting in a linear fashion, perhaps along its
seam, with the other PF pairs splitting apart somewhat behind this leading
crack-like defect. Indeed, materials do typically split along linear cracks,
where the elastic stresses are concentrated \cite{Kanninen1985}. In particular,
splitting between random PFs would yield step sizes that, due to helicity, could
be a small fraction of the tubulin size. It would be hard to physically motivate
a range $\delta$ that is more than ten times the incremental depolymerization
step size. Why would such a depolymerization process, involving little or no
motion of PFs more than $1$~nm from the last fully polymerized section of MT, be
highly sensitive to the presence of a \DASH ring more than $10$~nm distant? We
therefore consider our estimate of the characteristic range $\delta$ for the
burnt-bridges reaction (within which the \DASH ring occludes unzipping) to be
quite reasonable.

\begin{figure}
\begin{center}
\includegraphics[width=3.25in]{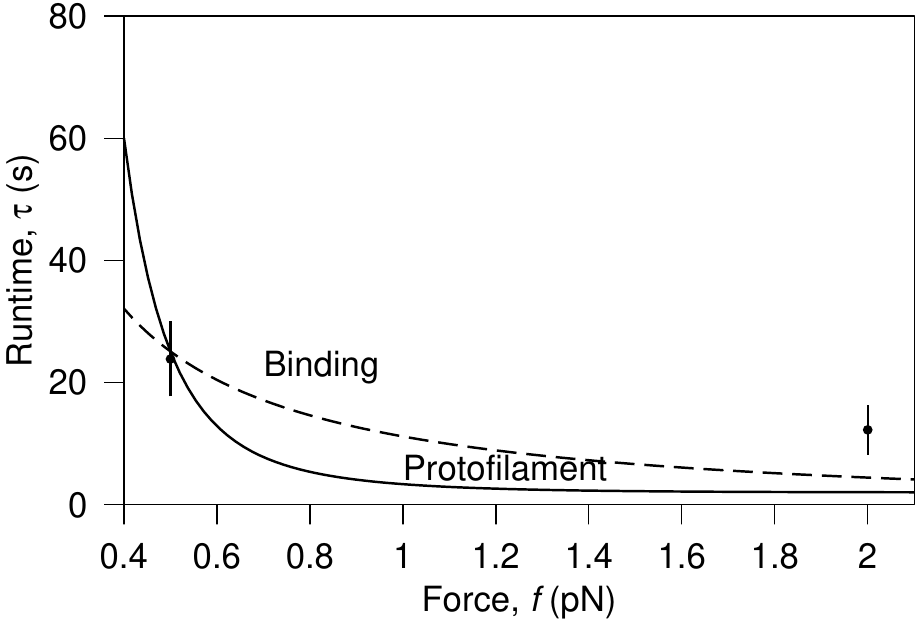}
\caption{Runtime of the \nh and \nsb models. The runtime $\tau$ of each model is
  calculated using the parameters fitted as described in the Results
  section. Although, it may seem that distinguishing the models by varying force
  is possible due to the difference between their predicted behaviour, as shown
  here, the difference is close to experimental error ($\pm6.15$ s) and both
  models present similar functional form. Only two data points with sufficient
  statistics were available to perform this fitting \cite{FrancPGGDA2007} making
  it difficult to draw any conclusions from this approach. The fit provides
  values for $\mtdam$ for the \nsb model and $\kb$ for the \nh model.}
\label{fig:runtime}
\end{center}
\end{figure}

Combining the available data for velocity and detachment frequency we find, on
average, $\res=23.9$~s and $12.2$~s, for $f=0.5$~pN and $2.0$~pN
respectively. To fit the \nsb model for $\res$ we choose $\epsilon=1$~nm, as a
reasonable distance over which an attraction might act, and find
$\mtdam=15\pm0.26$~\kt. Independently, we fit $\kb$ for the \nh model and find
$\kb=7.1\pm0.63$~$\text{s}^{-1}$. Fitting these parameters to just two data
points does not provide strong evidence for these particular values. However,
uncertainty in the exact parameter values should not detract from the main value
of this work; to provide a model that explains the \DASH force sensitivity and
to distinguish between \nsb and \nh models. Comparison of the fit with data is
shown in \figref{fig:runtime}.

\begin{figure*}
\begin{center} 
\includegraphics[width=6.5in]{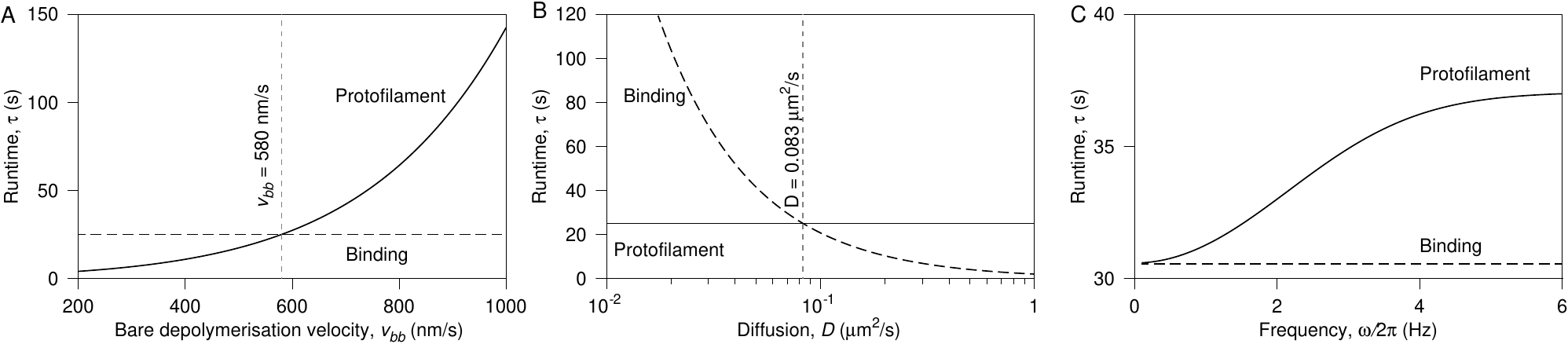}
\caption{Model discrimination. The panels show variation of runtime $\res$ with
  ({\it A}) bare MT depolymerization velocity $\vbb$, ({\it B}) diffusion
  coefficient $D$, and ({\it C}) frequency of applied force $\omega/2\pi$, for
  both models under load $f=0.45$~pN, chosen because both models predict the
  same nominal $\res$ and $v$ at this load (see \figref{fig:runtime}). ({\it A})
  The runtime $\res$ increases exponentially with $\vbb$ for the \nh model,
  whilst the \nsb model is insensitive. This is because the \nh model directly
  depends on $v$, but the \nsb model does not. ({\it B}) Restricted diffusion
  suppresses detachment for the \nsb model because $\res$ is inversely related
  to $D$, due to the reduced impetus to escape the potential barrier. The \nh
  model, on the other hand, is not affected by $D$ since $\tzip$ is independent
  of $D$. Distinguishing between models will be easiest by experimental
  reduction of $D$, for example by attachment of a long polymer. ({\it C}) The
  \nsb model is sensitive only to the amplitudes $f_0$ (here 0.1~pN) and $f_1$
  (here 0.43~pN), not the frequency $\omega$. The rate of detachment for the \nh
  model instead strongly depends on the frequency: roughly speaking the ring is
  lost more quickly when the high force part of the cycle persists for long
  enough for the PFs to completely depolymerize in this time, i.e. when the
  period is long.}
\label{fig:results}
\end{center}
\end{figure*}

\subsection*{Variation of intrinsic depolymerization velocity}

The \nh model exhibits the most sensitivity to the intrinsic (bare) MT
depolymerization velocity $\vbb$, as is shown in \figrefl{fig:results}{A}. For
the \nh model, \res\ is strongly dependent on $\mean{\tzip}$ and consequently
$\vbb$. The \nsb model, on the other hand, is only weakly dependent on $v$ (see
{\it Supporting Material}), and on this range of $\vbb$ we can assume that
depolymerization is quasistatically slow with respect to ring diffusion. The
result can be understood physically by realising that as $\vbb$ increases, the
rate of PF unzipping $\kz$ also increases, while $\kb$ remains constant making
it less likely that the PFs will break off sufficiently quickly to release the
ring.

An experimental test that might be able to distinguish which model operates
could be achieved, e.g., by addition of a depolymerization inducing agent, such
as $\text{Ca}^{2+}$ or XMCAK1.

\subsection*{Changing of diffusion coefficient}

The diffusion constant $D$ of the ring is determined by the ring's dimensions
and the roughness of the binding energy landscape along the MT, rather than the
magnitude of the binding energy itself. A more rough landscape reduces the
mobility of the ring. \figrefl{fig:results}{B} shows the effect of the diffusion
constant on the runtime for both models. Only the \nsb model is sensitive to
change in $D$, having reduced runtime with faster diffusion. This is because the
increased mobility of the ring increases the chance it is able to scale the
potential barrier constraining it to the MT.

Although it may be possible to alter $D$ biochemically, for example by
phosphorylation \cite{GestaGCWZWAD2008}, it is difficult to do so independently
of $\mtdam$. Decreasing $D$ may be better accomplished by attaching a long inert
polymer to the complex to increase viscous drag.

\subsection*{Effect of time-varying loading force}

The runtime in the \nsb model is sensitive only to the instantaneous force  provided $2\pi/\omega \gg \lambda^2/D$, see the low frequency portion of \figrefl{fig:results}{C}. If $f_1=1$~pN then $2\pi/\omega_{\text{max}}$ is on the order of 1~kHz. The runtime in the \nh model is sensitive to the time over which changes in $v$ persist. If the force is oscillating with a long period then the rate of detachment will be greater in the high force part of the cycle  than if the period is short. This is because the \DASH ring takes some time to detach if it needs to first wait for the PF curls to break, see \figrefl{fig:results}{C}. Sigmoidal increase of $\res$ would be a signature of a system that depends on a second time (1/$\kb$), like the \nh model; insensitivity of $\res$ to frequency would  imply a \nsb-style coupling.

\section*{Conclusion}

Our results indicate a power stroke does contribute to the effective force
generated during depolymerization but only becomes dominant at over 2~pN
load. We show how a faster depolymerization mechanism must operate at lower
loads and argue that the \DASH ring suppresses depolymerization when it is close
to the MT end.

We have shown that either of two rather different \DASH-MT coupling mechanisms
might be operating under piconewton loads. Both models have comparable
performance under load; their differences only become apparent under novel
experimental conditions. Structural studies cannot resolve the question of which
model operates {\it in vivo}. We suggest several methods for using runtime
statistics to determine which class of model best describes the coupling of the
\DASH ring to depolymerizing MTs.  Note that throughout this study we have
assumed that depolymerization is sufficiently slow compared to ring diffusion
that we can consider the distribution of the ring's position to be
quasi-equilibrated. Over the range of parameters we have considered this
assumption is valid to within $1\%$ of the predicted velocity.  The
characteristic range over which Dam1 inhibits depolymerization can be estimated
by comparing our model with data. It is intriguingly close to the size of the
microscopic (tubulin) repeat length of the MT. We have argued that this provides
evidence that the MT is splitting, possibly along its seam, at the leading edge
of the depolymerization front. In this case the PFs move outwards a similar
distance along the MT from the last polymerized section.

It is important to note that the present work has neglected {\it in vivo}
factors such as microtubule-associated proteins (MAPs) or kinases. However, some
of these factors operate to increase or reduce the depolymerization rate of the
microtubule, a parameter included in the model. We therefore expect the general
results to remain largely applicable. Furthermore, we have assumed Dam1 to be
present as a ring. Recent work \cite{GestaGCWZWAD2008,GrishSVEWDBAM2008} has
raised the possibility that Dam1 may operate as short oligomers or single
complexes. If we can assume these oligomers interact with PFs in a comparable
fashion as a ring would, our model would be indistinguishable for rings or
oligomers. If not, our model may be of use to determine whether ring or oligomer
is present based on e.g., differing diffusion constant.

\section*{Acknowledgements}
The authors thank George Rowlands and Jonathan Millar for useful discussions.

\bibliographystyle{biophysj}
\bibliography{paper_arxiv}
\end{document}